\documentclass[10pt]{book}

\usepackage[s5paper,laser]{antero}
\usepackage{cite}
\usepackage{amsfonts}
\usepackage{amssymb}
\usepackage{latexsym}
\usepackage{bbm}

\newtheorem{theo}{Theorem}

\relax
\catcode`\@=11
\def\defop#1{\expandafter\def\csname#1\endcsname{\mathop{\rm#1}\nolimits}}
\def\deflie#1{\expandafter\def\csname#1\endcsname{\mathop{\mathfrak{#1}}\nolimits}}
\let\@@line\|                           

\def\<#1>{\left\langle#1\right\rangle}	
\def\norm#1{\mathopen\@@line#1\mathclose\@@line}

\let\@AA=\AA				
\def\AA{\ifmmode\hbox{\rm\@AA}\else\@AA\fi} 
\def\slash{\@ifnextchar[{\@slash}{\@slash[\z@]}}
\def\@slash[#1]#2{\setbox\z@\hbox{$#2$}\@tempdima\wd\z@\box\z@%
\@tempdimb#1 \advance\@tempdimb-\@tempdima \kern\@tempdimb
\hbox to\@tempdima{\hss\@makeslash\hss}}
\def\@makeslash{$/$}			
\catcode`\@=12

\relax
\def\d{{\rm d}} 
\defop{Hom}  \defop{Aut}   \defop{End}   \defop{Diff} \defop{Perm}
\defop{Map}  \defop{Mat}   
\defop{GL}   \defop{SL}    \defop{O}     \defop{SO}   
\defop{SU}   \defop{Sp}    \defop{Spin}  \defop{Pin}   
\defop{Cl}   \defop{Cliff} \defop{Osp}  \defop{OSp}
\defop{tr}   \defop{Tr}    \defop{Det}   \defop{Det}   
\defop{Pf}   \defop{pf}	   \defop{str}   \defop{Str}   
\defop{sdet} \defop{Sdet} 
\defop{diag} \defop{rank}  \defop{codim} \defop{Index}
\defop{ad}   \defop{Ad}    \defop{Id}    \defop{id} 
\defop{Re}   \defop{Im}    \defop{sgn}   \defop{sgn} 
\defop{im}   \defop{span}  \defop{sign}  \defop{Ch} 
\defop{Lie} 

\deflie{gl}   \deflie{o}   \deflie{so}    \deflie{u}
\deflie{su}   \deflie{sp} \deflie{cliff}  \deflie{cl}	
\deflie{spin} \deflie{pin} \defop{osp}    \deflie{g}
\newcommand{\be}{\begin{equation} }
\newcommand{\ee}{\end{equation} }
\newcommand{\ba}{\begin{eqnarray} }
\newcommand{\ea}{\end{eqnarray} }

\begin{document}

\renewcommand{\thepage}{\roman{page}}\setcounter{page}{5}

\begin{titlepage}

\vskip 0.7truecm

\begin{center}
{ \bf ON PATH INTEGRAL LOCALIZATION AND THE LAPLACIAN, THE THESIS \\  }
\end{center}

\vskip 1.5cm

\begin{center}
{\bf Topi K\"arki $^{*}$ } \\
\vskip 0.4cm
{\it Institute Mittag-Leffler, 
Aurav\"agen 17,\\ S-182\,62, Djursholm, Sweden}
\vskip 0.4cm
\end{center}

\rm
\noindent

\vspace*{10mm}

\noindent

In this thesis, we develop path integral localization methods that are familiar from topological field theory: the integral over the infinite dimensional
integration domain depends only on local data around some
finite dimensional subdomain.

We introduce a new localization principle that unifies BRST localization, the non-Abelian localization principle and the conformal generalization
of the Duistermaat-Heckman integration formula. 

In addition, it is studied if one can possibly derive a generalized
Selberg's trace formula on locally homogeneous manifolds using localization
techniques. However, a definite answer is obtained only in the Lie group case (we complete the work of R. Picken) in which it is an application of the Duistermaat-Heckman integration formula. Also a new derivation of DeWitt's term
is reported. 

Furthermore, connections between evolution operators of integrable models and
localization methods are studied. A derivative expansion localization
is presented and it is conjectured to apply also to integrable models, for example the Toda lattice.

Moreover, a pedagogical introduction to the localization techniques is given, as well as a list of selected references that might be useful for a beginning graduate student in mathematical physics or for a mathematician who would like to study the physical point of view to topological field theory and
string theory.

\vfill

\vfill
31.15.K, 02.40.H, 02.20.T
\begin{flushleft}
\rule{5.1 in}{.007 in}\\
$^{*}$ {\small E-mail: Topi.Karki@Helsinki.Fi \\}
\end{flushleft}
\end{titlepage}
\newpage

\clearpage
\thispagestyle{empty}

\noindent This thesis is based on the following papers

\bigskip

\begin{list}{-}{\renewcommand{\makelabel}[1]{\makebox[\labelwidth][l]{#1}}}

\item[I.]
T. K\"arki and A. Niemi, {On the Duistermaat-Heckman integration
formula and integrable models}, Ahrenshoop Symp. 1993, 175; hep-th/9402041.

\item[II.] 

T. K\"arki and A. Niemi, {\em Supersymmetric quantum mechanics and the De Witt
effective action}, Phys.~Rev.~D. {\bf 56}, 2080 (1997).
\item[III.] 

T. K\"arki, {\em On path integral localization
and the Laplacian}, J.~Math.~Phys. {\bf 40}, 1807 (1999); hep-th/9712169.
\end{list}

\bigskip

\noindent Reprints were made with permissions from the publishers.

\renewcommand{\thepage}{\arabic{page}} \setcounter{page}{1}
\tableofcontents

\chapter{Introduction}
Ever since I. Newton invented differential
calculus simultaneously with G. Leibniz the development
of theoretical physics and mathematics have been closely related.

In the recent decades this interplay has manifested, for instance,
in integrable models, topological field theories and string theory. Such abstract mathematics as Donaldson theory, Floer cohomology, Jones polynomial, Calabi-Yau manifolds, representation theory of infinite dimensional
Lie groups, intertwining operators, modular forms, non-commutative geometry, quantum groups etc. have become familiar to physicists.
This is hardly the end, the spectacular development
of the second string revolution has
given hope that all the five string theories can be unified
in a single M theory in 11 dimensions (not to mention
D-branes and black hole physics). When it is achieved,
it is probably going to effect mathematics in a profound way as well as our understanding
of space-time\footnote{We thank
C. Montonen for pointing this out.}, of which we have had some flavour already in the form
of non-commutative geometry.
For the interested reader we have collected in Appendix~\ref{appa} some references that we have found useful for further reading.

A very central object in todays mathematical physics is the partition function
that is formulated as Feynman's 
path integral. Developing methods to calculate it is of great importance
because it may enrich our understanding of non-perturbative
quantum field theories. A lot is already understood:
topological field theories, conformal field theories, integrable models, supersymmetric field theories
and various dualities provide exact results and intriguing conjectures.
The research in this thesis has concentrated on one particular phenomenon:
path integral localization.
It is familiar from topological field theories,
the integral over the infinite dimensional integration domain
actually depends only on local data around some finite dimensional subdomain.
More specifically, we have been interested in if this method has applications
beyond the apparently 'trivial' topological field theories. As a testing
ground we have chosen the most trivial case, namely the Laplacian on locally
homogeneous manifolds,
that should not be overlooked as it often happens that mathematical structures
repeat themselves in the most unexpected places. It is reasonable
that understanding of this specific model might help one to understand
also gauged WZW models that involve the most obvious generalization of
Lie groups, the affine groups (or perhaps alternatively $\sigma$-models,
our recent unpublished research suggests however rather WZW models).
Not to mention that developing the localization method is important
in its own right, we conjecture that in the long run it might increase our understanding
of evolution operators of integrable models.

This thesis is organized as follows: In Ch. \ref{brst} we give an elementary introduction to the BRST symmetry
which is the basis for localization. In Ch. \ref{llit} we illustrate
the localization technique with the general Laplacian that serves also as a toy model of topological field theory.
In Ch. \ref{dhll} we introduce the model of our main interest, the Laplacian on functions,
and comment on the semiclassical exactness of it on compact Lie groups, or in other words
on the Duistermaat-Heckman integration formula in the infinite dimensional setting.  

Then we advise the reader to papers I-III hoping that the preceding chapters have provided a pedagogical introduction.
The goal of the research has been to derive generalized Selberg's
trace formulae on locally homogeneous manifolds using physicist's
path integral. They are of interest for both mathematicians
and physicists because they can be used in
calculating high loop string amplitudes that result
in determinants of the Laplacian on Riemann surfaces. 
The goal is not achieved but some other interesting results
are obtained: A new derivation of DeWitt's term, a generalized canonical transformation that unifies previously unrelated localizations, new formulas on homogeneous spaces that might be useful in a hypothetical Selberg's trace formula localization, a non-trivial localization deformation on homogeneous spaces,
a derivative expansion localization that might also be useful
for integrable models and a contribution
to the problem 'Why is the semiclassical approximation exact for the heat 
kernel on Lie groups?'. The last contribution is clarified in Ch. \ref{dhll} of this thesis.

\chapter{BRST symmetry in gauge theories}\label{brst}
The BRST symmetry is the basis for path integral localization. We review\footnote{Our exposition follows closely L. Baulieu's unpublished lecture notes.} it shortly in
the historical context where it was discovered, in Yang-Mills theories:
We begin with the Faddeev-Popov procedure.
The partition function for four dimensional Yang-Mills can be written as
\begin{equation}
Z=\int [dA]\delta ({\cal G^A})\det\left|\frac{\delta{\cal G}}{\delta\omega}
\right|\exp S_c.
\end{equation}
${\cal G}^A$ is the gauge condition, eg. ${\cal G}^B=\partial_{\mu}A^{B\mu}$ is the Lorentz gauge,
and 
\begin{equation} 
S_c=\int d^4x\,\mbox{tr}\, F_{\mu\nu}F^{\mu\nu}\label{caction}
\end{equation}
is the Yang-Mills action\cite{ramond} for some gauge group $G$.
The differentiation $\frac{\delta{\cal G}}{\delta\omega}$ requires some further explanation,
the variation in the numerator actually means the effect of an
infinitesimal gauge variation on the gauge condition ${\cal G}$
that is parametrized by $\omega^B$. More explicitly,
$\delta A_{\mu}^B=D_{\mu}\omega^B$ so that for the Lorentz gauge
$\frac{\delta{\cal G}}{\delta\omega}=\partial_{\mu}D^{\mu}$, where
$D_{\mu}=\partial_{\mu}+A_{\mu}$ is the covariant derivative.

The Faddeev-Popov procedure introduces a functional Fourier transform
of the delta function\footnote{Compare $\delta(x)\sim\int_{-\infty}^{\infty}e^{ipx}dp$.}  
\begin{equation}
\delta({\cal G})=\int[db]\exp \int b^A{\cal G^A}
\end{equation}
and a ghost path integral representation of the determinant
\begin{equation}
\det\left|\frac{\delta{\cal G}}{\delta\omega}\right|=\int[dc\,d\bar{c}]
\exp\int\bar{c}\frac{\delta{\cal G}}{\delta \omega}c
\end{equation}
where $c^B,\, \bar{c}^B$ are Grassmann fields.
In the Lorentz gauge the resulting path integral is
\begin{equation}
Z=\int[dA\,db\,dc\,d\bar{c}]\exp \int \mbox{tr}\,F_{\mu\nu}^2+b\partial\cdot A
+\bar{c}(\partial\cdot D)c,\label{qaction}
\end{equation}
we denote the action in (\ref{qaction}) as $S_q$ in contrast to the
action in (\ref{caction}).
It possesses a BRST symmetry $Q$, $QS_q=0$, which is a supersymmetry (it exchanges bosonic
and fermionic fields) and satisfies $Q^2=0$.
Its action on the fields is
\begin{eqnarray}
QA_{\mu}^B&=&D_{\mu}c^B\\
Qc^A&=&-\frac{1}{2}f^{ABC}c^Bc^C\\
Q\bar{c}^A&=&b^A\\
Qb^A&=&0
\end{eqnarray}
and it is a graded derivation (ie. like the exterior derivative
on forms) by the grading of the ghost number,
$b$ and $A$ are even while the ghosts $c$, $\bar{c}$ are odd.

$S_q$ can be written as the Yang-Mills action $S_c$ plus a BRST exact term
\begin{equation}
S_q=S_c+Q\psi
\end{equation}
where 
\begin{equation}
\psi=\bar{c}^B\partial\cdot A^B.\label{gaugefermion}
\end{equation}

The Batalin-Vilkovisky theorem states that the partition function
(\ref{qaction}) is independent of the gauge fermion $\psi,$
different $\psi$'s correspond to different gauge fixings. It can be proved
as follows \cite{henneaux}: Consider the partition function
\begin{equation}
Z=\int[d \Phi ]\exp S[\Phi]+Q\psi\label{PhiZ}
\end{equation}
where $\Phi$ means collectively all the fields $(A,\,b,\,c,\,\bar{c})$.
An infinitesimal change of variables ($\delta\psi$ is an infinitesimal
gauge fermion ie. infinitesimal real variable times
a gauge fermion) 
\begin{equation}
\delta \Phi=\delta \psi\, Q\Phi
\end{equation}
gives a super-Jacobian $\exp Q\delta\psi$ and the action does not
change because of the BRST symmetry $QS=0$. Thus, $\psi$ in equation (\ref{PhiZ})
is replaced by $\psi+\delta\psi$ and therefore
the partition function (\ref{PhiZ}) is independent of $\psi.\Box$

For example, if one considers instead the gauge fermion
\begin{equation}
\psi_{\mbox{standard}}=\bar{c}^B({\cal G}^B +\alpha b^B),
\end{equation}
one gets after the integration of the auxiliary $b$ field
$$S=\int \mbox{tr}\,F_{\mu\nu}^2+\frac{1}{2\alpha}{\cal G}^A{\cal G}^A+
\bar{c}(\frac{\delta{\cal G}}{\delta \omega})c$$
which is one of the standard gauge fixings.

\chapter{Localization, the Laplacian and index theorems }\label{llit}
We illustrate path integral localization with the Laplacian,
some excellent reviews are Refs. \cite{tqftphysrep} and
\cite{2qcmoore}. Despite the apparent
triviality, this model shows many of the features of topological
field theories (Donaldson theory, topological $\sigma$-models,
BF-theories etc.). 

On a Riemannian manifold one can write the general Laplacian
on forms as
\begin{equation}
\triangle=dd^*+d^*d=[d,d^*]_{\mbox{graded}}.
\end{equation}
The $Z_2$ grading of the exterior algebra is $$\bigwedge M=\left(\bigoplus_{k=0,2\ldots}
\bigwedge^kM\right)\oplus\left(\bigoplus_{k=1,3\ldots}\bigwedge^kM\right)=\wedge_{+}M\oplus\wedge_{-}M,$$ and it induces also a grading for the linear operators
that map $\bigwedge M$ onto itself. 
The exterior derivative $d$ and its dual $d^*$ are odd (or fermionic) operators because
they map even forms to odd forms and vice versa.

The partition function in operator formalism is 
\begin{equation}
Z=\mbox{Str}\, \mbox{e}^{-\beta [d, d^*]},\label{oppart}
\end{equation}
where $\mbox{Str}=\mbox{tr}(-1)^N$ and $N$ is the operator that gives the degree
of the form, eg. for a p-form $\omega$, $N\omega=p\omega.$

The path integral presentation of the partition function (\ref{oppart})
is
\begin{equation}
Z=\int[\sqrt{g}\,dx^{\mu}\,d\psi^{\mu}\,d\bar{\psi}_{\mu}]\exp\int_0^{\beta} g_{\mu\nu}\dot{x}^{\mu}\dot{x}^{\nu}+\bar{\psi}_{\mu}\nabla_t\psi^{\mu}+R^{\mu\nu}_{\kappa\lambda}\bar{\psi}_{\mu}\bar{\psi}_{\nu}\psi^{\kappa}\psi^{\lambda}\label{papart}
\end{equation}
where one has periodic boundary conditions for both the bosons $x^{\mu}$
and the fermions $\bar{\psi}_{\mu},\, \psi^{\mu}.$
The action in (\ref{papart}) results from canonical quantization \cite{dewitt},
\begin{equation}
S=\int p_{\mu}\dot{x}^{\mu} +\bar{\psi}_{\mu}\dot{\psi}^{\mu}-\{d,d^*\}\label{canq}
\end{equation}
which after integration of the auxiliary field $p$ gives exactly (\ref{papart}).
In equation (\ref{canq}) $d,\, d^*$ are the classical limits
of the corresponding operators, ie. functions on the super phase space, and $ \{    \:    ,   \:   \} $
is the graded Poisson bracket
\begin{eqnarray}
\{ x^{\mu},\,p_{\nu} \}&=&\delta^{\mu}_{\nu}\\
\{\psi^{\mu},\bar{\psi}_{\nu}\}&=&\delta^{\mu}_{\nu}
\end{eqnarray}
that is  associated with the kinetic term $\int p_{\mu}\dot{x}^{\mu}+\bar{\psi}_{\mu}\dot{\psi}^{\mu}.$ In the following we use also the loop space version of it which
is obtained as follows: if $$\omega=\omega_{\mu\nu}d\phi^{\mu}\wedge d\phi^{\nu}$$
is the corresponding symplectic form on the phase space $\Gamma$ with local coordinates
$\phi^{\mu},$  then the loop space symplectic form is
$$\omega=\int_0^{\beta}\omega_{\mu\nu}\delta\phi^{\mu}\wedge\delta\phi^{\nu}.$$

The action in (\ref{canq}) can be written as a BRST exact variation
$$S=Q(\int\dot{x}^{\mu}\bar{\psi}_{\mu}-d^*)$$
where the BRST operator $Q=\{d,\,\}$
satisfies
\begin{eqnarray}
Qx^{\mu}&=&\{ d,x^{\mu} \} =-\psi^{\mu}\\
Qp_{\mu}&=&0\\
Q\psi^{\mu}&=&0\\
Q\bar{\psi}_{\mu}&=&p_{\mu}.
\end{eqnarray}
One can argue that because the BRST operator does not
depend on the metric and because
of the Batalin-Vilkovisky theorem, the partition function is independent
of the metric. Thus, it is a topological
invariant. In the Hamiltonian formalism it is even more evident:
The Hamiltonian operator $\hat{H}=\triangle$ satisfies
$$[d,\hat{H}]=0$$
so that we can choose the exterior derivative $d$ to be the BRST operator $Q$.
The Batalin-Vilkovisky theorem in the operator formalism
can be stated and proved as follows:
\begin{theo}
Provided that the Hamiltonian $\hat{H}$ has a BRST symmetry $Q$, ie.
fermionic operator that satisfies 
$$[Q,\hat{H}]=0,\,\, [Q,Q]=0,$$
the partition function
\begin{equation}
Z=\mbox{Str}\, \mbox{e}^{-\beta \hat{H}+[Q,\psi]}
\end{equation}
is independent of the odd operator (gauge fermion) $\psi$.
\end{theo}
{\it Proof.}
We define 
$$Z_{\lambda}=\mbox{Str}\, \mbox{e}^{-\beta \hat{H}+[Q,\psi_{\lambda}]},$$
where $\psi_{\lambda}$ is a one-parameter family of gauge fermions.
Then 
\begin{eqnarray}
\frac{\partial Z_{\lambda}}{\partial\lambda}&=&\mbox{Str}\, [Q,\frac{\partial\psi_{\lambda}}{\partial\lambda}]\,
\mbox{e}^{-\beta\hat{H}+[Q,\psi_{\lambda}]}\\
&=&\mbox{Str}\,[Q,\frac{\partial\psi_{\lambda}}{\partial\lambda}\mbox{e}^{-\beta\hat{H}+[Q,\psi_{\lambda}]}]
\end{eqnarray}
which vanishes because the supertrace of a graded commutator is zero.
It is easy to see that any two gauge fermions can be connected by a one-parameter
family of such fermions.$\Box$

In particular we see that the partition function (\ref{oppart}),
or equivalently (\ref{papart}), is independent of $\beta$ because
it can be absorbed in the gauge fermion. In the limit $\beta\rightarrow\infty$ 
it gives
$$Z=\sum^D_{i=0}(-1)^i\,\mbox{dim}\,\mbox{ker}\,\triangle_i=\chi(M)
$$
where $\triangle_i$ is the Laplacian on i-forms.
On the other hand, evaluating the partition function in the limit $\beta\rightarrow 0$ gives the heat kernel proof of the Gauss-Bonet index theorem \cite{bergezver}.
The result is an integral over the Euler class of the tangent bundle,
\begin{equation}
Z=\int_{M}\mbox{Pf}\, R,
\end{equation}
where $R$ is the curvature tensor.
It can be seen most easily in the path integral
form (\ref{papart}). The limit $\beta\rightarrow 0$ corresponds to the dimensional reduction: all the fields $(x^{\mu},\psi^{\mu},\bar{\psi}_{\mu})$ are considered
to be independent of time, which results in a zero dimensional
theory on the zero modes:
\begin{eqnarray}
Z&=&\beta^{-D/2}\int\sqrt{g}\,dx^{\mu}d\psi^{\mu}d\bar{\psi}_{\mu}\,\mbox{e}^{\beta R^{\mu\nu}_{\kappa\lambda}\bar{\psi}_{\mu}\bar{\psi}_{\nu}\psi^{\kappa}\psi^{\lambda}}\label{X}\\
&=&\int_{M}\mbox{Pf}\, R.
\end{eqnarray}

But, as we proved in Theorem 1, the operator $d^*$ can actually be
replaced by any odd operator $\psi$, which shows that
one has a lot more freedom in evaluating the partition
function.
We consider another way of evaluating it which results in the Poincar\'e-Hopf
theorem (or more generally the Matthai-Quillen theorem).
We use E. Witten's Morse theoretic twist \cite{susymorse}
\begin{equation}
d\rightarrow d_s=\mbox{e}^{sf}d\mbox{e}^{-sf},\,\,\,\, d^*\rightarrow d^*_s=\mbox{e}^{-sf}
d^*e^{sf},
\end{equation}
(where it is assumed that $f$ has only
isolated critical points and that the Hessian is nondegenerate) and consider the generalized partition function
\begin{equation}
Z=\mbox{Str} \, \mbox{e}^{-\beta [d_s, d^*_s]}.
\end{equation}
It is independent of $s$ and $\beta$
because it can be written as
$$\mbox{tr}\,\,\mbox{e}^{sf}\left( (-1)^N\mbox{e}^{-\beta[d,d^*_{2s}]}\right)
\mbox{e}^{-sf}=\mbox{Str}\,\mbox{e}^{-\beta [d, d^*_{2s}]}$$
and because of Theorem 1.
When $s$ is zero it coincides with the partition function
(\ref{oppart}).
In the path integral form
\begin{eqnarray}
Z=\int[\sqrt{g}\,dx^{\mu}\,d\psi^{\mu}\,d\bar{\psi}_{\mu}]\exp \,-\int_0^{\beta} g_{\mu\nu}\dot{x}^{\mu}\dot{x}^{\nu}+\bar{\psi}_{\mu}\nabla_t\psi^{\mu}+R^{\mu\nu}_{\kappa\lambda}\bar{\psi}_{\mu}\bar{\psi}_{\nu}\psi^{\kappa}\psi^{\lambda}\\
+s\bar{\psi}_{\mu}g^{\mu\kappa}\frac{D^2f}{Dx^{\kappa}Dx^{\nu}}\psi^{\nu}
+s^2g^{\mu\nu}\partial_{\mu}f\partial_{\nu}f\nonumber
\end{eqnarray}
which in the limit $\beta\rightarrow 0$ gives using the dimensional
reduction (see Appendix \ref{appb} for more details) 
\begin{eqnarray}
Z=\beta^{-D/2}\int\sqrt{g}\,dx\,d\psi\,d\bar{\psi}\,\exp -\beta (R^{\mu\nu}_{\kappa\lambda}
\bar{\psi}_{\mu}\bar{\psi}_{\nu}\psi^{\kappa}\psi^{\lambda}\label{Y}\\
+s\bar{\psi}_{\mu}g^{\mu\kappa}\frac{D^2f}{Dx^{\kappa}Dx^{\nu}}\psi^{\nu}
+s^2g^{\mu\nu}\partial_{\mu}f\partial_{\nu}f)\nonumber
\end{eqnarray}
which is called the Matthai-Quillen form. In the limit
$s\rightarrow\infty$ the integral localizes around
the critical points $df=0.$
Thus, it is legitimate to expand in Taylor series around the critical points
and sum the contributions
(it amounts to using the delta function formula $\delta(x)\sim\sqrt{\lambda}\exp -\lambda x^2$). For example $g^{\mu\nu}\partial_{\mu}f\partial_{\nu}f=g^{\mu\nu}\partial_{\mu\kappa}f\partial_{\nu\lambda}fx^{\kappa}x^{\lambda}$ where the coordinate system is chosen so that $x=0$ at the particular critical point whose contribution is calculated.

The resulting integrals are Gaussian in the limit and can be summed to give
\begin{equation}
Z=\sum_{df=0}\mbox{sign} \,\det \left(\frac{\partial^2 f}{\partial x^{\mu}
\partial x^{\nu}}\right)=\sum_{df(p)=0}(-1)^{n_p},
\end{equation}
which is the Poincare-Hopf theorem. The integer
$n_p$ denotes the morse index at point $p$, ie. the number of negative eigenvalues of the Hessian.

Concluding, we have showed that using mixed path integral and
operator techniques it is relatively easy to derive Gauss-Bonnet and
Poincare-Hopf index theorems. The methods that we have described can be used
to derive many if not all the index theorems, see Refs. \cite{alvarezgaumez} and \cite{bergezver}. This is even more interesting
because they can be understood as toy models of topological quantum field theory
(see the references in Appendix \ref{appa}). For example, similarly some correlation
functions of Donaldson theory, which is a twisted $N=2$ supersymmetric
Yang-Mills theory, can be localized on the moduli space of instantons
giving Donaldson polynomials. Finally, for the reader who is interested in the
Laplacian we mention that the partition function of the BF theories
is essentially the Ray-Singer torsion \cite{tqftphysrep}
$$T_M=\prod_{k=0}^D(\det \triangle_k)^{(-1)^k k/2}$$
(where $\triangle_k$ is the Laplacian on k-forms) that is a topological invariant.

\chapter{DH theorem and the Laplacian on Lie groups}\label{dhll}

The model of main interest in this thesis is the Laplacian
on zero-forms $\triangle_0.$
The partition function of it is in the path integral form
on a general Riemannian manifold\footnote{We neglect DeWitt's term (paper II),
it can be thought of as being part of the measure.}
\begin{equation}
Z=\mbox{tr}\, \mbox{e}^{-\beta \triangle_0}=\int [\sqrt{g}\,dx] \mbox{e}^{
\int_0^{\beta} g_{\mu\nu}\dot{x}^{\mu}\dot{x}^{\nu}}.\label{star}
\end{equation}

The model is particularly interesting 
because it is known to possess localization behaviour:
it is semiclassically exact on Lie groups and on Riemann surfaces it
can be evaluated using Selberg's trace formula (see paper III).
We comment on the former clarifying our result in paper III and completing
the seminal work of R. F. Picken in Ref. \cite{pic}.

On a compact Lie group $G$ the heat kernel can be written
as a path integral over the
paths in $G$ satisfying the boundary conditions $g(0)={\bf 1},\,\,
g(\beta)=g$:

\begin{eqnarray}
k_{\beta}({\bf 1}, g)&=&\int[dg]\exp \int \mbox{tr}(g^{-1}\partial_t g)^2\\
&=&\int[\sqrt{\det(K_{ij}\omega^i_{\mu}\omega^j_{\nu})}\,dx^{\mu}]\exp\int K_{ij}\omega^i(\dot{x})\omega^j(\dot{x})\label{2stars}
\end{eqnarray}
where $$T_i\omega^i_{\mu}dx^{\mu}=g^{-1}dg$$ are the left invariant vector fields and $T_i$
are the generators of the Lie algebra having the Killing form
$$K_{ij}=-\mbox{tr}\,T_iT_j=-<T_i, T_j>.$$
On the Lie group one has the unique bi-invariant metric
$$g_{\mu\nu}=K_{ij}\omega^i_{\mu}\omega^j_{\nu}.$$

The integral (\ref{2stars}) can be transformed into an integral over
the based loop space \cite{pic} using the change of variables
\begin{equation}
g(t)\rightarrow g_0(t)g(t)
\end{equation}
where $g_0(t)$ is a geodesic connecting ${\bf 1}$ and $g,$
\begin{equation}
g_0(t)=\mbox{e}^{(J^iT_i)\frac{t}{\beta}}
\end{equation}
where $J^i$ are constants.
The boundary conditions in the integral (\ref{2stars}) change to
periodic and the action to
\begin{equation}
S=\int_0^{\beta}K_{ij}(\omega^i(\dot{x})+J^i)(\omega^i(\dot{x})+J^i).\label{Sham}
\end{equation}

The loop space can be interpreted as an
infinite dimensional flag manifold which in particular means that
it is a K\"ahler manifold. From the localization
point of view it is important that it possesses a right invariant symplectic form which at the unity is given by the Maurer-Cartan cocycle \cite{pic}
\begin{equation}
\omega(X,Y)=\int_0^{\beta}K_{ij}X^i\partial_tY^j.\label{tahti}
\end{equation}
$X=X^i(t)T_i\in L\mbox{g}$, $X^i(0)=0$, is an element of the 
Lie algebra of the based loop space.

Using right translations one obtains the symplectic form
\begin{equation}
\omega=\int_0^{\beta}K_{ij}\omega^i_R\partial_t\omega^j_R
\end{equation}
where $dg\,g^{-1}=T_i\omega^i_R$ are the right invariant one-forms. This can be seen easily
by noticing that it is annihilated by the Lie derivatives ${\cal L}_
{\int J^i(t)v_i}$ ($v_i$ are
the left invariant vector fields that generate the right action) and that it coincides at the unit element with (\ref{tahti}). 

The Liouville measure associated with this symplectic form
equals the natural bi-invariant measure $[\sqrt{g}\,dx]$ in equation
(\ref{2stars}), as has been argued heuristically in paper III,
so that one can finally write
\begin{equation}
k_{\beta}({\bf 1}, g)= \int [\sqrt{\omega}\,dx]\exp \int K_{ij}(\omega^i(\dot{x})+J^i)(\omega^j(\dot{x})+J^j)\label{2tahti}
\end{equation}
where the integral is over the based loop space.
It has been shown in Ref. \cite{pic} that an application of the Duistermaat-Heckman integration
formula
\begin{equation}
Z=\sum_{\delta S=0}\frac{\sqrt{\det \omega_{\mu\nu}}}{\sqrt{\frac{\delta^2 S}{
\delta x^{\mu}\delta x^{\nu}}}}\mbox{e}^S\label{DHformula}
\end{equation}
yields the exact expression for the heat kernel, which, as the measure
coincides with the metric one, amounts to the semiclassical exactness
of the theory.

We comment briefly on the path integral level of rigour why the Duistermaat-Heckman theorem can be applied. The point is that the Hamiltonian
(\ref{Sham}) generates a torus action\footnote{We thank O. Tirkkonen
for conjecturing this and the following comment in the paranthesis.} (probably it is even a projective
Hamiltonian on the flag manifold in the sense of Ref. \cite{pic})
which allows one to use an invariant metric and the BRST localization
principle.

We denote 
\begin{eqnarray}
H&=&\int_0^{\beta}K_{ij}\omega^i(\dot{x})\omega^j(\dot{x})\\
I_i&=&\int_0^{\beta}K_{ij}\omega^j(\dot{x}).
\end{eqnarray}
The corresponding Hamiltonian vector fields are
\begin{eqnarray}
\tilde{\chi}&=&\dot{x}-\left.\omega^i(\dot{x})\right|_{t=0}v_i^R\\
u_i&=&v_i-v_i^R.
\end{eqnarray}
The Poisson brackets of the Hamiltonians read
\begin{eqnarray}
\{H,I_i\}&=&0\\
\{I_i,I_j\}&=&C^k_{ij}I_k.
\end{eqnarray}

The flow associated with $\tilde{\chi}$ is just the rotation of the loops
\begin{equation}
g(t)\rightarrow g(\alpha)^{-1}g(t+\alpha)
\end{equation}
and the vector field $u_i$ generates the adjoint action 
\begin{equation}
g(t)\rightarrow h(\gamma)g(t)h(\gamma)^{-1}
\end{equation}
where $h(\gamma)=\mbox{e}^{\gamma T_i}$ \cite{pressley}.

Altogether, we have a Hamiltonian action of the group $U(1)\times G$
on the phase space of based loops $\Omega G$.
We suppose in addition that $J^iT_i$ is in the generic position that the
elements of the Lie algebra that annihilate it generate a Cartan
subalgebra $H$ \cite{fomenko}, we denote by $T$ the associated
maximal torus.  Then the torus $U(1)\times T$ has a Hamiltonian action
on the phase space and in particular the Hamiltonian
\begin{equation}
S=\int K_{ij}(\omega^i(\dot{x})+J^i)(\omega^j(\dot{x})+J^j)=
H+J^i I_i+ \mbox{constant}\label{haayoaie}
\end{equation}
generates an element of the torus group.
We construct an invariant metric $g$ by averaging an arbitrary metric
$\tilde{g}$ over the action of the torus\footnote{It is enough to average over the group $U(1)\times G$ because $G$ is compact, however, we would like to emphasize the appearance of a torus action.}
\begin{equation}
g=\int_{\vec{t}\in U(1)\times T}d\vec{t}\, \varphi_{\vec{t}}^*\tilde{g}
\end{equation}
where $\varphi_{\vec{t}}$ is the diffeomorphism associated with the element
$\vec{t}$ of the torus.
Then the Hamiltonian vector field $\chi$ associated to the action (\ref{haayoaie}) satisfies the Lie derivative condition
\begin{equation}
{\cal L}_{\chi}g=0
\end{equation}
and we are in position to apply the BRST proof of the Duistermaat-Heckman
theorem (paper I).

We exponentiate the symplectic measure by introducing a Grassman integral
\begin{equation}
k_{\beta}({\bf 1}, g)= \int[dx^{\mu}\d\psi^{\mu}]\mbox{e}^{S+\omega}
\end{equation}
and using the BRST symmetry 
\begin{equation}
(d+i_{\chi})(S+\omega)=0,
\end{equation}
where $i_{\chi}$ is the contraction operator by the vector field
$\chi$, we add the gauge fermion
$$\psi=\lambda i_{\chi}g$$
obtaining
$$k_{\beta}({\bf 1},g)=\int[dx^{\mu}d\psi^{\mu}]\exp \left(S+\omega
-\lambda g(\chi,\chi)-\lambda d(i_{\chi}g)\right)$$
which in the limit $\lambda\rightarrow \infty$ localizes
on the zeros of $\chi$ giving the Duistermaat-Heckman formula
(\ref{DHformula}).

In paper III we have constructed explicitly an invariant
tensor $g'$ which localizes equally well, if it would be
nondegenerate the treatment there would be exactly parallel to
the one given here. And finally, we end the discussion by mentioning that
the semiclassical exactness holds also on the noncompact groups
\cite{mar2}, but it seems to be difficult
to explain it using similar path integral arguments.

\cleardoublepage
\def\rightmark{Summary of the papers}
\def\leftmark{Summary of the papers}
\addcontentsline{toc}{chapter}{Summary of the papers}
\vspace*{15mm}
\noindent
{\Huge\bf Summary of the papers}

\vspace{10mm}

\begin{list}{-}{\renewcommand{\makelabel}[1]{\makebox[\labelwidth][l]{#1}}}

\item[{\bf I.}]
The integral proof of the Duistermaat-Heckman theorem
is reviewed. Connections between the Duistermaat-Heckman theorem
and integrable models are speculated. A geodesic condition
for the Hamiltonian vector field is reported
as a possible generalization of the isometry condition appearing
in the DH theorem. A bi-Hamiltonian
structure associated with the new condition is presented and
it is speculated if it underlies a new localization principle.

\item[{\bf II.}]
A new derivation of DeWitt's term is presented.
It results from considering the grand canonical partition function
of the supersymmetric quantum mechanics associated with
the Laplacian on general forms. The Laplacian on zero-forms
together with DeWitt's term
appears in the limit that the chemical potential
is put to infinity. The numerical value of DeWitt's term
seems to be connected with the cancellation of quantum mechanical anomalies. 
\item[{\bf III.}] 
A new localization principle is reported. It unifies
BRST localization, non-Abelian localization principle and
the conformal DH formula of Paniak, Semenoff and Szabo.
It is also applied to the Laplacian on homogeneous manifolds
but it fails to give the speculated localization that would
generalize Selberg's trace formula. A
derivative expansion is used to localize the Laplacian on Lie groups and
it is conjectured to apply also to integrable models.
In addition the semiclassical exactness on Lie groups
is explained by completing R. F. Picken's work using a path integral
proof. 

\end{list}

\cleardoublepage
\def\rightmark{Acknowledgements}
\def\leftmark{Acknowledgements}\addcontentsline{toc}{chapter}{Acknowledgements}
\vspace*{15mm}
\noindent
{\Huge\bf Acknowledgements}

\vspace{10mm}

\noindent
I thank prof. A. Niemi for the opportunity to learn research in
such an interesting branch of mathematical physics.
I have learned so much from him and his research group. I thank all the other members of the research group: J. Kalkkinen, M. Miettinen, O. Tirkkonen,
K. Palo, E. Keski-Vakkuri and P. Pasanen. To this list of people I add
M. Laine, J. Lukkarinen and P. Stjernberg. The Appendix \ref{appa} of this
thesis would not exist without these people. 

I owe a special thank to O. Tirkkonen who contributed to my work
by his conjectures and by showing me some very fundamental papers.
I thank also M. Laine for help with a technical (but very important)
problem in paper II and J. Kalkkinen in paper III.

I thank all the senior scientists who have offered their precious time
and discussed with me. In particular, I thank A. Kupiainen, A. Alekseev
and A. Losev who helped me to detect a mistake in the early version
of paper III.

I thank Helsinki institute of physics for funding most of the work
but also Uppsala university and Institut Mittag-Leffler for hospitality.
The financial support of Suomen kulttuurirahasto and T. and M. Veuro is
gratefully acknowledged.
In addition I thank
A. Tuure for providing me a cheap computer that was critical
for this thesis, both in research and in writing it.
I thank also M. M\"akel\"a for teaching me how to use it and for
all the trouble he had in installing Linux on it.

\appendix

\chapter{Technical details}\label{appb}
\def\leftmark{Appendix \ref{appb}. ~Technical details}
We explain in a little more detail the dimensional reduction:

We consider the partition function
\begin{eqnarray}
Z&=&\int [dx\,dp\,d\psi\,\d\bar{\psi}] \exp \int_0^{\beta}p_{\mu}\dot{x}^{\mu}
+\bar{\psi}_{\mu}\dot{\psi^{\mu}} -H_s\\
&=& \int [dx\,dp\,d\psi\,\d\bar{\psi}] \exp \int_0^{1}p_{\mu}\dot{x}^{\mu}
+\bar{\psi}_{\mu}\dot{\psi^{\mu}} -\beta H_s,
\end{eqnarray}
where $H_s=\{ d_s,\,d^*_s \}$, and use periodic boundary conditions
for all the fields. The limit $\beta\rightarrow 0$ can be evaluated
by scaling time dependent modes $x_t$ ($x^{\mu}=x^{\mu}_0+x^{\mu}_t,$
where $x_0$ is the constant Fourier mode and $x_t$ the rest
of the expansion). More explicitly, we use the change of variables 
\begin{eqnarray}
x_t \rightarrow  \sqrt{\beta}x_t,\,\,\, p_t \rightarrow  \sqrt{\beta}p_t\\
\psi_t \rightarrow  \sqrt{\beta}\psi_t,\,\,\, \bar{\psi}_t \rightarrow \sqrt{\beta}\bar{\psi}_t
\end{eqnarray}
and because the measure splits as $$dx_0\,dp_0\,d\psi_0\,d\bar{\psi}_0
[dx_t\,dp_t\,d\psi_t\,d\bar{\psi}_t]$$
the Jacobian is $1$.
The action becomes
\begin{equation}
S=\beta\int_0^1p_{t\mu}\dot{x}^{\mu}_t+\bar{\psi}_{t\mu}\dot{\psi}^{\mu}_t-H(x_0,p_0,\psi_0,\bar{\psi}_0) +O(\beta^{1/2})
\end{equation}
and in the limit $\beta\rightarrow 0$ one can evaluate the integral over the time dependent modes 
\begin{equation}
\int dx_t d\psi_t\delta(\dot{x}_t)\delta(\dot{\psi}_t)=\sqrt{\frac{\det\delta^{\mu}_{\nu}\partial_t}{\det \delta^{\mu}_{\nu}\partial_t}}
=1
\end{equation}
so that the final result is
\begin{equation}
\int dx_0\,dp_0\,d\psi_0\,d\bar{\psi}_0 \,\exp -\beta H_s
\end{equation}
where $\beta$ is small (actually the result is equivalent to the classical
limit $\hbar\rightarrow 0$ which is not so surprising because
$\beta$ and $\hbar$ appear in almost the same way). However, because
of the BRST structure $H_s=\{d_s,d_s^*\}$ the result is independent of $\beta$
so that it can be considered to be finite instead. 
In Eqs. (\ref{X}), (\ref{Y}) the factor $\sqrt{g}\beta^{-D/2}$ results from integrating the
$dp_0$ integral.

\chapter{Selected references}
\label{appa}
\def\leftmark{Appendix \ref{appa}. ~Selected references}
\section{Field theory}\label{sec:field}
\def\rightmark{\ref{sec:field}. ~Field theory}
Basic references on quantum field theory are \cite{ramond}, \cite{coleman},
\cite{weinberg}, \cite{blove}, \cite{skyrme} and \cite{nakahara}.
In order to understand renormalization and for many other reasons
it is good to have some idea of lattice field theory \cite{montway}, critical phenomena in statistical physics and Wilson's renormalization group. For the
two latter we unfortunately do not know any particularly good reference (except perhaps
conformal field theory on which we comment later).
Some information on solitons, monopoles and instantons
can be found in Refs. \cite{rajaraman}, \cite{harvey}, \cite{harvey2},
\cite{atiyah} and \cite{nashsen}, for deeper understanding we suggest
integrable models and duality in supersymmetric gauge theories on which we comment later.
Some references on constrained quantization and BRST are Ch. \ref{brst} of this
thesis and Refs. \cite{baulieu}, \cite{henneaux}, \cite{henneaux2}, \cite{gomis}. Topological field theory provides perhaps the best place to see it in action,
see the section on toy models below.
Important background for grand unified theories and almost
all theoretical physics are Lie groups, the best reference for physicists
can be found in one chapter of the book \cite{francesco}.
One book on grand unified theories is \cite{gut} and a very
short and easy introduction can be found in Ref. \cite{halzen}.
We also recommend to take a look at the end of the volume two of the
superstring theory book \cite{gsw}.
The Fujikawa method for anomalies is described in Ref. \cite{nakahara},
see also \cite{weinberg} and \cite{gsw} for the Feynman graph point of view.
But we recommend to study string theory for deeper understanding, see the
references in string theory section below.
For confinement see Refs. \cite{confinement}, \cite{t'hooft} and the following
section on supersymmetry.

However, the best way to learn field theory is to study string theory,
which, despite the apparently naive idea, leads to a surprisingly
realistic theory. Many recent advances in field theory have
been derived from string theory in the low energy field theory
limit.

\section{Supersymmetry}\label{sec:susy}
\def\rightmark{\ref{sec:susy}. ~Supersymmetry}
Supersymmetry, although not observed in the nature, is of great importance
in theoretical physics.

Good basic references are \cite{blove}, \cite{lykken},
\cite{west}, \cite{sohnius}, \cite{kksugra}.
The fastest way to learn supersymmetry is to study
the exact solution of $N=2$ supersymmetric Yang-Mills theory that was discovered by Seiberg and
Witten. One of the most pedagogical references is  \cite{bilal},
but there are also a lot of other lecture notes on the internet \cite{othersw} not
to mention the original papers \cite{sw}.

The importance of Seiberg-Witten theory is difficult to
exaggerate \cite{today}. It has proved that t'Hoofts monopole condensation mechanism \cite{t'hooft} actually occurs in certain supersymmetric theories, and because
Donaldson theory is a twisted $N=2$ theory it has had in addition
remarkable consequences for topology. Furthermore this
duality has been extended to string theories leading to the
discovery of M theory. We give some references on these other developments below.

\section{String theory}\label{sec:string}
\def\rightmark{\ref{sec:string}. ~String theory}
And finally, we come to the most important section of this appendix: String theory.
The most pedagogical references are perhaps the books
\cite{gsw}, \cite{blove}, \cite{polchinsky} together with the unpublished
proceedings \cite{trieste} and the thesis \cite{thesis}.
For string theory compactifications (see also Ref. \cite{appelquist} where the original Kaluza-Klein compactification is presented: 5 dimensional gravity
gives an effective theory of gravity and electrodynamics in the Minkowski space) one needs also background in algebraic geometry, for which we recommend Refs. \cite{gsw}, \cite{hubsch} and \cite{alggeom}.
The discovery of D-branes and dualities have revolutionized the subject and
despite Polchinski's book \cite{polchinsky}, and perhaps the proceedings \cite{tasi} and the thesis \cite{thesis}, the information is scattered on
the internet (hep-th bulletin board: xxx.lanl.gov; slac-spires conference and proceedings info: www.slac.stanford.edu/find/spires.html, especially
Nordita and Trieste conferences; 
various internet sites which gather reviews on the web; various institutes
that have also online proceedings,...).
 
We give some references for these more recent developments, but apologize for
possible misleading comments as we are not really specialists in string theory:

The basic idea of describing branes as strings with Dirichlet boundary
conditions (ie. branes are objects on which the end points of
open strings must move) is explained in \cite{notesond}. A very pedagogical
treatment can be found in Ref. \cite{billo}. In the low energy field theory
they are brane-like solitons, see Refs. \cite{stringsolitons} and \cite{stelle}.
The string theory dualities are treated pedagogically in Ref. \cite{townsend}
but also E. Witten's original paper \cite{witdua} is very pedagogical.
But there are also a lot of other lectures on the subject on the internet, for example Refs. \cite{otherstringduality}. The string dualities lead also to the conjecture about M theory
which in the low energy limit should be the eleven dimensional supergravity. It is the only supergravity
in dimension 11 and there are no supergravities in higher dimensions (if we forget F theory on which we do not comment at all, although it may be very interesting
and important). The matrix model conjecture associated with M theory is described pedagogically in Refs. \cite{polchinsky} and \cite{bilalmatrix}.

The unexpected bonus of the recent developments has been the increased
understanding of ordinary field theories. On $p$-branes one has the
effective field theory which is the D=10, N=1 super Yang-Mills
dimensionally reduced to $p+1$ dimensions, the scalars of the theory
describe the fluctuations of the brane.
The $SU(N)$ gauge theory in $p+1$ dimensions can be described as putting $N$ Dp-branes on the top of each other \cite{branesfields}.
A very important recent development is the Maldacena conjecture
which relates the large $N$ structure of the $N=4$ theory with 
type IIB string theory \cite{witmal}. It confirms the long speculated
relation of large $N$ gauge theory and string theory \cite{coleman}.

The most intriguing 'recent' triumph of string theory
has been the black hole entropy calculation of C. Vafa using D-branes \cite{blackholes}, it seems that string theory is finally approaching its original goal:
understanding of quantum gravity.

From the mathematical point of view string theory is equally interesting: vertex operator algebras, modular forms, mirror symmetry (T-duality), Calabi-Yau manifolds, superconformal field theory and non-commutative geometry have found their place in physics.
And there is probably a lot more to be discovered as the
final formulation of M theory is still lacking.  

\section{Toy models}\label{sec:toy}
Some insights into field theory and string theory can be obtained by studying different toy models, which are also very important in their own right.
A very quick introduction to many of the subjects in this section can be found
in Ref. \cite{nash}.

Conformal field theory in two dimensions is important because it is the mathematical structure
underlying string theory and because it is solvable (minimal or rational CFTs) it provides many insights into quantum
field theory. 
Particularly good references are \cite{petersen}, \cite{beletal}, \cite{francesco} and \cite{polchinsky}. The mathematicians point of view can be found in Ref. \cite{segal}. We recommend the reader
to study in particular the cosets models that are the gauged WZW models (rational CFTs) and have the enlarged affine group symmetry.

Even more interesting is superconformal field theory because its relations
to mirror symmetry, quantum cohomology, topological sigma models
on Calabi-Yau manifolds etc. Unfortunately our knowledge ends here. 

Integrable models are another source of intuition to quantum field theory, in particular to soliton physics, see
Refs. \cite{takhtajan}, \cite{faddeev}, \cite{fadtak} , \cite{bogolubov} and
\cite{perelomov}. 
Integrable models are the integration tables of our time and they repeat themselves in all branches of physics, notably in topological field theories like
two dimensional gravity.
The field is so large that the reference list given here is
very subjective.

Good references on topological field theory are \cite{witten-TYM}, \cite{tqftphysrep}, \cite{2qcmoore} and \cite{Dijkgraaf}. See Ref. \cite{kauffman} for background on knots. We recommend the reader also to take a look at
the papers of M. Blau and G. Thompson on the hep-th bulletin board.
For Seiberg-Witten invariants we recommend the references in the
previous supersymmetry section and to search the internet for lectures, eg.
Refs. \cite{variousrefs}.
A particularly interesting model of topological field theory is Chern-Simons theory, which
has also a connection to three dimensional gravity. One might even gain understanding of black holes by studying such
toy models.

\def\rightmark{Bibliography}
\def\leftmark{Bibliography}


\begin{thebibliography}{II}
\addcontentsline{toc}{chapter}{Bibliography}
\frenchspacing
\def\rightmark{Bibliography}
\def\leftmark{Bibliography}



\bibitem{ramond}
P. Ramond, {\em Field theory: a modern primer}
(Addison-Wesley, 1990).
\bibitem{henneaux}M. Henneaux, {\em
Hamiltonian form of the path integral for theories with a gauge
freedom,}
Phys. Rept. {\bf 126}, 1 (1985). 
\bibitem{tqftphysrep}
D.~Birmingham, M.~Blau, M.~Rakowski and G.~Thompson, {\em Topological field theory,} Phys.~Rept.~{\bf 209}, 129 (1991).
\bibitem{2qcmoore}
Stefan Cordes, Gregory Moore, Sanjaye Ramgoolam 
{\em Lectures on 2-D Yang-Mills theory, equivariant cohomology and
topological field theories,} Nucl. Phys. Proc. Suppl. {\bf 41}, 184 (1995) (Part 1 of the lectures is published in the
Nucl. Phys. B, Proc. Suppl. and part 2 in the Les Houches proceedings. The entire set of lectures is
posted as hep-th/9411210.), also in Trieste Spring School 1994, p. 184. 
\bibitem{dewitt}
T. K\"arki and A. Niemi, {\em Supersymmetric quantum mechanics and the De Witt
effective action,} Phys. Rev. D {\bf 56}, 2080 (1997).
\bibitem{bergezver}
 N. Berline, E. Getzler and M. Vergne, {\em Heat kernels and Dirac operators}
(Springer-Verlag, 1992).
\bibitem{susymorse}
E. Witten, {\em Supersymmetry and Morse Theory,} J. Diff. Geom. {\bf 17}, 661 (1982). 
\bibitem{alvarezgaumez}L. Alvarez-Gaume,
{\em Supersymmetry and the Atiyah-Singer index theorem,}
Commun. Math. Phys. {\bf 90}, 161 (1983).
\bibitem{pic}  R. Picken, {\em Loop spaces and a geometrical approach to path integral
quantization,} Contributed to 11th Workshop on Geometric Methods in Physics, Bialowieza, Poland, 7-14, in Bialowieza 1992, Proceedings, Quantization and coherent states methods, p. 75,\\
{\em The propagator for quantum mechanics on a group manifold from
an infinite dimensional analog of the Duistermaat-Heckman
integration formula,} J. Phys. A {\bf 22}, 2285 (1989),\\
{\em The Duistermaat-Heckman integration formula on flag
manifolds,} J. Math. Phys. {\bf 31}, 616 (1990). 
\bibitem{pressley}A. Pressley and G. Segal, {\em Loop groups}
(Oxford Univ. Press, 1986).
\bibitem{fomenko}
A. Fomenko, {\em Differential geometry and topology} (Consultants Bureau, 1987).
\bibitem{mar2}  N. Krausz and M. Marinov, {\em  Exact evolution operator on non-compact group manifolds,} quant-ph/9709050.
\bibitem{coleman}
S. Coleman,
   {\em Aspects of symmetry: selected Erice lectures of Sidney
   Coleman} (Cambridge Univ. Press, 1985).
\bibitem{weinberg} 
S. Weinberg, {\em The quantum theory of fields} (Cambridge Univ. Press, 1995). 2v.
\bibitem{blove}D. Bailin and A. Love,
   {\em Supersymmetric gauge gield theory and string theory} (IOP, 1994).
\bibitem{skyrme}
I. Zahed and G.E. Brown, {\em The Skyrme model}, Phys. Rept. {\bf 142}, 1 (1986).
\bibitem{nakahara} M. Nakahara,
   {\em Geometry, topology and physics} (Hilger, 1990).
\bibitem{montway}I. Montvay and G. Muenster,
   {\em Quantum fields on a lattice} (Cambridge Univ. Press, 1994).
\bibitem{rajaraman}
 R. Rajaraman, {\em Solitons and instantons: an introduction to solitons and
   instantons in quantum field theory}  (North-Holland,
   1982).
\bibitem{harvey}J. Harvey, {\em Magnetic monopoles, duality and supersymmetry,}
 Trieste HEP Cosmology 1995, 66; hep-th/9603086. 
\bibitem{harvey2} C. Callan, J. Harvey and A. Strominger,
{\em Supersymmetric string solitons,}
Trieste 1991, Proceedings, String theory and quantum gravity '91, 208; hep-th/9112030. 
\bibitem{atiyah}
M. Atiyah and N. Hitchin,
{\em The geometry and dynamics of magnetic monopoles}
(Princeton Univ. Press, 1988).
\bibitem{nashsen} C. Nash and S. Sen, {\em Topology and geometry for physicists} (Academic Press Limited, 1989).

\bibitem{baulieu}
L. Baulieu, {\em Perturbative gauge theories,}
Phys. Rept. {\bf 129}, 1 (1985).
\bibitem{henneaux2}
M. Henneaux and C. Teitelboam, {\em Quantization of gauge systems}
(Princeton Univ. Press, 1992).
\bibitem{gomis}
J. Gomis, J. Paris and S. Samuel, {\em Antibracket, antifields and gauge theory quantization,} Phys. Rept. {\bf 259} 1 (1995); hep-th/9412228. 

\bibitem{francesco} P. Di Francesco, P. Mathieu and D. Senechal,
{\em Conformal field theory} (Springer-Verlag, 1997).
\bibitem{gut} G. Ross, {\em Grand unified theories}
(Benjamin/Cummings, 1984).
\bibitem{halzen}F. Halzen and A. Martin,
{\em Quarks and leptons: an introductory course in modern particle
   physics} (Wiley, 1984).
\bibitem{gsw}M. Green, J. Schwartz and E. Witten,
   {\em Superstring theory} (Cambridge Univ. Press, 1987).
   2v.
\bibitem{confinement}
M. Bander, {\em Theories of quark confinement,}
Phys. Rept. {\bf 75}, 205 (1981).
\bibitem{t'hooft}G. t'Hooft, {\em A property of electric and magnetic flux in nonabelian gauge
theories,}
Nucl. Phys. B {\bf 153}, 141 (1979), \\
{\em Topology of the gauge condition and new confinement phases in
nonabelian gauge theories,} Nucl. Phys. B {\bf 190}, 455 (1981).  
\bibitem{lykken}J. Lykken,
{\em Introduction to supersymmetry,}
Talk given at Theoretical Advanced Study Institute in Elementary Particle Physics (TASI 96), hep-th/9612114. 
\bibitem{west}
P. West, {\em Introduction to supersymmetry and supergravity} (World Scientific, 1990). 2nd ed.
\bibitem{sohnius}
M.~Sohnius, {\em Introducing supersymmetry,}
Phys.~Rept.~{\bf 128}, 39  (1985).
\bibitem{kksugra}
M. Duff, B. Nilsson and C. Pope, {\em Kaluza-Klein supergravity,}
Phys. Rept. {\bf 130}, 1 (1986). 
\bibitem{bilal}A. Bilal,
{\em Duality in N=2 SUSY SU(2) Yang-Mills theory: a pedagogical
introduction to the work of Seiberg and Witten,} hep-th/9601007. 
\bibitem{othersw}
M. Shifman: hep-th/9704114, C. Gomez: hep-th/9510023, 
L. Alvarez-Gaume and S. Hassan: hep-th/9701069, 
L. Alvarez-Gaume and F. Zamora: hep-th/9709180,
P. Di Vecchia: hep-th/9803026.
\bibitem{sw} N. Seiberg and E. Witten,
{\em Monopoles, Duality and Chiral Symmetry Breaking in N=2 Supersymmetric QCD,}
Nucl. Phys. B {\bf 431}, 484 (1994); hep-th/9408099,\\
{\em Monopole Condensation, And Confinement In $N=2$ Supersymmetric Yang-Mills Theory,} Nucl. Phys. B {\bf 426}, 19 (1994); Erratum-ibid. B {\bf 430}, 485 (1994); hep-th/9407087.
\bibitem{today}G. Collins, {\em Supersymmetric QCD sheds light on quark confinement and the topology of 4-manifolds,}
Physics Today, March 1995, p. 17.
\bibitem{polchinsky}
J. Polchinski, {\em String theory} (Cambridge Univ. Press 1998). 2v.
\bibitem{trieste}
Introductory School on String Theory (Strings 98): 8-19 Jun 1998, Trieste, Italy, unpublished proceedings (we thank secretary J. Varnier
for kindly mailing them to us).
\bibitem{thesis}
A. Westerberg, {\em Strings, branes and symmetries,} Ph. D. thesis,
Institute of theoretical physics, Chalmers university of technology and
G\"oteborg university, G\"oteborg 1997;
http://fy.chalmers.se/~tfeawg/.
\bibitem{appelquist}T. Appelquist, A. Chodos and P. Freund,
   {\em Modern Kaluza-Klein theories } (Benjamin/Cummings, 1985).
\bibitem{hubsch} T. H\"ubsch, {\em Calabi-Yau manifolds : a bestiary for physicists} (World Scientific, 1992).
\bibitem{alggeom}P. Griffiths, {\em Introduction to algebraic curves} 
(American Mathematical Society, 1989).\\P. Griffiths and J. Harris, {Principles of algebraic geometry}
(Wiley, 1994).
\bibitem{tasi}
Theoretical Advanced Study Institute in elementary particle
physics (TASI 96): Fields, strings, and duality,\\
Proceedings: C. Efthimiou and B. Greene, {\em TASI 96, Fields, strings and duality,} (World Scientific, 1997)\\
or see http://www.slac.stanford.edu/find/spires.html, 'conferences' and after
searching the conference
'List of Papers submitted to the Meeting/Conference'.
\bibitem{notesond}
J. Polchinski, S. Chaudchuri and C. Johnson, {\em Notes on D-branes,}
hep-th/9602052.\\
J. Polchinski, {\em TASI lectures on D-branes,} hep-th/9611050. 
\bibitem{billo} M. Billo, {\em Introduction to D-branes,} Nordita May 96
unpublished proceedings.
\bibitem{stringsolitons}
M. Duff, R. Khuri and J. Lu, {\em String solitons,}
Phys. Rept. {\bf 259}, 213 (1995); hep-th/9412184.
\bibitem{stelle}K. Stelle, {\em BPS branes in supergravity,} 
Talk given at ICTP Summer School in High-energy Physics and Cosmology, Trieste, Italy, 10 Jun - 26 Jul 1996 and at the ICTP Summer School in High-Energy Physics and Cosmology Trieste, Italy, 2 Jun - 11 Jul 1997. 
In Trieste 1997, High energy physics and cosmology p. 29; hep-th/9803116. 

\bibitem{townsend}
P. Townsend, {\em Four lectures on M theory,} 
Talk given at Summer School in High-energy Physics and Cosmology, Trieste, Italy, 10 Jun - 26 Jul 1996. 
In Trieste 1996, High energy physics and cosmology, p. 385;
hep-th/9612121.  
\bibitem{witdua}
E. Witten, {\em 
String theory dynamics in various dimensions,} 
Nucl. Phys. B {\bf 443}, 85 (1995); hep-th/9503124.
\bibitem{otherstringduality}
C. Vafa: hep-th/9702201, E. Kiritsis: hep-th/9708130, A. Sen: hep-th/9802051,
J. Duff: hep-th/9608117, W. Taylor: hep-th/9801182.
\bibitem{bilalmatrix}A. Bilal, {\em M(atrix) theory: a pedagogical introduction,}
Fortsch. Phys. {\bf 47}, 5 (1999); hep-th/9710136.
\bibitem{branesfields}
A. Giveon and D. Kutasov, {\em Brane dynamics and gauge theory,}
hep-th/9802067. 

\bibitem{witmal}
E. Witten, {\em  Anti-De Sitter space and holography,}
Adv. Theor. Math. Phys. {\bf 2}, 253 (1998); hep-th/9802150.
\bibitem{blackholes} J. Maldacena,
{\em Black holes in string theory,} Ph.D. Thesis,
Princeton Univ.; hep-th/9607235,\\
{\em Black holes and D-branes,}
Lectures given at 33rd Karpacz Winter School of Theoretical Physics: Duality - Strings and Fields,
Karpacz, Poland, p. 13, Feb 1997. 
Published in Nucl. Phys. Proc. Suppl. {\bf 61A}, 111 (1998), Nucl. Phys. Proc. Suppl. {\bf 62}, 428 (1998); hep-th/9705078.  

\bibitem{nash}
C. Nash, {\em Differential topology and quantum field theory,}
(Academic Press 1991)\\
{\em Topology and physics: an historical essay,} 
hep-th/9709135.

\bibitem{petersen}J. Petersen, {\em Notes on conformal field theory,}
(notes prepared with the help of J. Rasmussen) unpublished lecture notes of a Nordita school.

\bibitem{beletal} 
A. Belavin, A. Polyakov and A.B. Zamolodchikov,
{\em Infinite conformal symmetry in two-dimensional quantum field theory,} 
Nucl. Phys. B {\bf 241}, 333 (1984) (also in C. Itzykson (ed.), et al.: {\em Conformal invariance
and applications to statistical mechanics}, p. 5, and in P. Goddard (ed.) and D. Olive (ed.):
{\em Kac-Moody and Virasoro algebras,} p. 413). 
\bibitem{segal} G. Segal, {\em The definition of conformal field theory,}
unpublished.
\bibitem{takhtajan} L. Takhtajan, {\em Lectures on quantum groups. Introduction to quantum group
and integrable massive models of quantum field theory,} (Nankai, 1989) Nankai Lectures Math. Phys., p. 69 (World Sci. Publishing, River Edge, NJ, 1990).
\bibitem{faddeev} L. Faddeev, {\em 
How algebraic Bethe ansatz works for integrable model,} hep-th/9605187,\\
{\em Algebraic aspects of Bethe ansatz,}
J. Mod. Phys. A {\em 10}, 1845 (1995); hep-th/9404013,\\
{\em Integrable models in (1+1)-dimensional quantum field theory,}
Les Houches Sum. School 1982, p. 561.
\bibitem{fadtak}
L. Faddeev and L. Takhtajan, {\em Hamiltonian methods in the theory of solitons}
(Springer-Verlag, 1987).
\bibitem{bogolubov}V. Korepin, N. Bogolyubov and A. Izergin, {\em Quantum inverse scattering method and correlation functions}
(Cambridge Univ. Press, 1993).
\bibitem{perelomov} A. Perelomov, {\em Integrable systems of classical mechanics and Lie algebras} (Birkhauser, 1990).\\
M. Olshanetskii and A. Perelomov, {\em Classical integrable finite dimensional systems related to Lie
algebras,} Phys. Rept. {\bf 71}, 313 (1981),\\
{\em Quantum integrable systems related to Lie algebras,}
Phys. Rept. {\bf 94}, 313 (1983). 

\bibitem{witten-TYM}
E.~Witten, {\em Topological quantum field theory,} Comm.~Math.~Phys.~{\bf 117}, 353 (1988),\\
{\em Quantum field theory and the Jones polynomial,}
Commun. Math. Phys. {\bf 121}, 351 (1989),\\
{\em Topological sigma models,}
Commun. Math. Phys. {\bf 118}, 411 (1988). 
\bibitem{Dijkgraaf}R. Dijkgraaf, {\em Les Houches Lectures on Fields, Strings and Duality,} hep-th/9703136.

\bibitem{kauffman}L. Kauffman, {\em Knots and physics} (World scientific, 1991).
\bibitem{variousrefs}E. Witten, {\em Monopoles and four manifolds,}
hep-th/9411102.\\
J. Labastida and C. Lozano, {\em Lectures in topological quantum field theory,} hep-th/9709192.\\ 
J. Labastida and M. Marino, {\em Duality and topological quantum field theory,}
hep-th/9704032. \\
G. Thompson, {\em New results in topological field theory and abelian gauge theory,} hep-th/9511038. 

\end{thebibliography}
\end{document}